# A deep learning framework for morphologic detail beyond the diffraction limit in infrared spectroscopic imaging

KIANOUSH FALAHKHEIRKHAH[1,2], KEVIN YEH[2], SHACHI MITTAL[2], LUKE PFISTER[2], ROHIT BHARGAVA[*,1,2,3,4,5,6]

[1] *Department of Chemical and Biomolecular Engineering, University of Illinois at Urbana- Champaign, Urbana, IL 61801*
[2] *Beckman Institute for Advanced Science and Technology, University of Illinois at Urbana- Champaign, Urbana, IL 61801*
[3] *Department of Bioengineering, University of Illinois at Urbana- Champaign, Urbana, IL 61801*
[4] *Department of Electrical and Computer Engineering, University of Illinois at Urbana- Champaign, Urbana, IL 61801*
[5] *Mechanical Science and Engineering, University of Illinois at Urbana- Champaign, Urbana, IL 61801*
[6] *Cancer Center at Illinois, University of Illinois at Urbana- Champaign, Urbana, IL 61801*
*\*rxb@illinois.edu*

**Abstract:** Infrared (IR) microscopes measure spectral information that quantifies molecular content to assign the identity of biomedical cells but lack the spatial quality of optical microscopy to appreciate morphologic features. Here, we propose a method to utilize the semantic information of cellular identity from IR imaging with the morphologic detail of pathology images in a deep learning-based approach to image super-resolution. Using Generative Adversarial Networks (GANs), we enhance the spatial detail in IR imaging beyond the diffraction limit while retaining their spectral contrast. This technique can be rapidly integrated with modern IR microscopes to provide a framework useful for routine pathology.

## 1. Introduction

Infrared (IR) spectroscopic imaging allows label-free and nondestructive recording of the chemical composition of samples. In conjunction with machine learning algorithms, this data can be used for assessment of cell type composition and diseases by measuring optical contrast in tissue[1–8] and by providing contrast mimicking conventional clinical images[9]. While IR imaging provides high quality spectral information due to its ability to access the characteristic vibrational modes of molecules, finer spatial detail is often lost compared to optical microscopy due to the 10-fold longer wavelengths and wider span of mid-IR wavelengths that restrict microscope design and performance. In general, the spatial quality of imaging systems operating in the IR spectrum is poorer than visible light microscopes commonly used in biomedical assessment[10]. Recent advances in instrumentation, particularly the combination of discrete frequency measurements[11–15] and high-definition optical design, have helped improve image quality considerably while optimizing the speed of data acquisition [16–18]. Ultimately, however, spatial detail is still limited compared to the commonly available imaging technologies used in pathology laboratories. More recent techniques that hybridize IR illumination with visible detection[19–23] or with nanoscale atomic force microscopy (AFM)[24,25] have provided better spatial resolution but do not yet provide spectral fidelity, speed and signal-to-noise ratio (SNR) that is comparable to that offered by direct absorption IR microscopy. These data characteristics are essential for biological recognition. Thus, there is a trade-off in the spatial and spectral domains that restrains progress. Here, we address this trade-off between image quality and spectral quality using a computational approach. Several attempts have been made to computationally enhance IR images using deconvolution[26,27], noise rejection[28,29] and other signal processing[30] approaches. Each of these approaches assumes a model for the optics, structure of noise in the data or property of the sample domains (e.g. smooth boundaries) but does not actually take advantage of an easily available resource, bright-field microscopy in routine pathology. This is the key advance in our approach here. We report a numerical framework that utilizes both high-resolution (HR) IR and visible images to learn the relative structure in the data, which can then be made capable of transforming a single band low-resolution (LR) IR data into computationally super-resolved (SR) high-resolution data sets.

Several different methods surpass the diffraction limit and achieve SR with far-field visible microscopy. These include popular techniques that take advantage of the non-linear response of fluorophores (e.g. STED[31]) as well as stochastic techniques that temporally separate emitters (e.g. PALM, STORM[32]). However, the ability to translate similar strategies to the IR spectrum is limited, though a framework exists[33], in a label-free manner. Computational approaches show more promise for estimating morphologic details, and three different categories of approaches have traditionally been offered to address this problem. The first are interpolation-based methods which create more pixels appearing to smooth and blur the result, but inherently add no additional information[34–36]. In the second category, the HR image is created from multiple LR images by using signal processing and computational techniques[37–39]. Lastly, recent developments in machine learning methods have demonstrated ability to estimate high spatial frequencies details when given an LR image; these convolutional neural networks (CNN) and their variants are a promising tool toward improving the performance of computational super-resolution[40–43].

Deep learning methods have been used to enhance the resolution in fluorescence and bright-field imaging[44–47] using a pre-trained CNN to map LR to HR images. While it is indeed possible to relate the two images, the process is ill-posed in the sense that several different high-resolution images may be consistent with the LR data. Additional information is often needed to correctly guide the recovery of the finer tissue textures and details. For example, this can be the semantic information of the HR objects. In IR spectroscopic imaging, the spectral profile of tissues is correlated to the chemical composition of the object. Since the spectra are conserved between the LR and HR images, being a property of the volume of material, this constraint inherently provides the additional information required. Based on this property, we developed a deep learning approach that is based on Generative Adversarial Networks (GAN)[48]. GAN-based methods use two models: a generator and a discriminator. The generator is a deep CNN used to learn the transformation between LR and HR images, while the discriminator critiques the generated image, returning feedback to the generator for iterative improvements toward minimizing a loss function. Using a large training dataset, complex network interconnections, and appropriate loss functions, GANs have been shown to approximate the complicated and nonlinear transformation between LR and HR data to produce realistic images[42,43]. However, it has been shown that the quality of GAN-generated images can be improved by incorporating additional semantic information, such as class labels. These so-called Conditional GANs (C-GAN) aim to learn the class-conditional image distribution. [49–52]. In the field of image super-resolution, where the generated images are an estimation, and completely different objects can have similar low spatial frequency details that appear identical, it is possible that the estimation converges incorrectly[53]. Therefore, in order to increase the accuracy of generated SR image, we hypothesize that C-GANs can prove useful. This technique also incorporates the class information of the objects in the image and enhances them accordingly[53]. In this study, we demonstrate the ability of C-GANs to increase the spatial resolution of a single band from a multispectral IR dataset. We show that C-GANs can improve the spatial resolution beyond the IR diffraction limit and we compare our results with unconditional GANs. Finally, we discuss the limitations of our presented approach and the potential challenges in its adoption.

## 2. Methods

### 2.1 Sample preparation and imaging

We imaged a breast tissue micro array (TMA) (BR1003, US Biomax Inc.) which is derived from formalin-fixed and paraffin-embedded (FFPE) tissue. The TMA consists of a total of 101 tissue cores from 47 patients; each core is 1 mm in diameter and is 5 μm thick. The tissue samples were processed using the same protocol typically used in preparing clinical samples. One section of the TMA was placed on a traditional glass slide, stained with Hematoxylin and Eosin (H&E), and subsequently imaged with a whole slide scanner. A serial section was placed on low-emissivity glass slide (IR reflective) for IR transflection imaging. The custom quantum cascade laser (QCL) discrete frequency IR (DF-IR) spectral imaging system[14] was used to measure absorbance images of the sample at 12 distinct wavenumber positions in the mid-IR fingerprint region. These frequencies and number are those typically required for classification[13]. The data was acquired at 2 μm spatial spacing with a 0.7 numerical aperture (NA) objective.

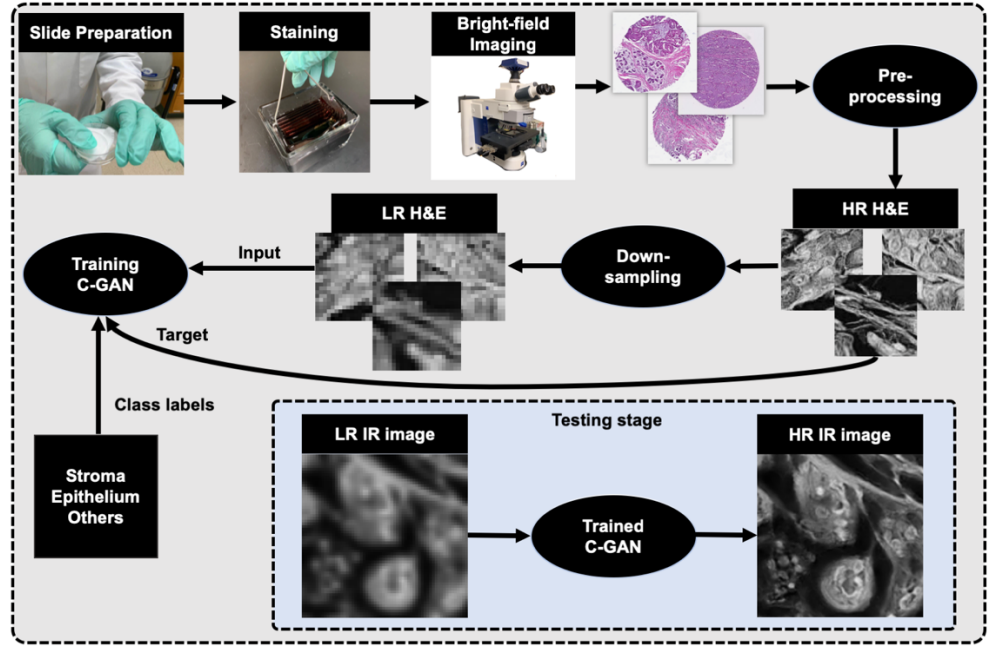

Figure 1. A super-resolution framework for increasing the spatial information of IR images based via deep learning. This framework has been trained on bright-field microscopic images and is capable of estimating IR images with a resolution beyond the diffraction limit.

### 2.2 Model design

Our deep learning framework utilizes the concept of C-GANs for enhancing image quality. We introduce a C-GAN framework (Fig. 1), which is designed to improve the spatial detail of IR images beyond its diffraction limit to match those from high resolution bright-field microscopy. First, H&E stained images are down sampled, preprocessed, and divided to smaller tiles to simulate the IR images in order to train the network. It is important to note that the IR images are not part of the training phase as the model only utilizes the stained data at that point. During training, the down-sampled and preprocessed H&E data is used as the input of the network along with semantic information representing the histological class for each tile: stroma, epithelium, and miscellaneous cell types bundled as "others". We also compared our C-GAN approach with an unconditional GAN (U-GAN) that does not incorporate any class labels.

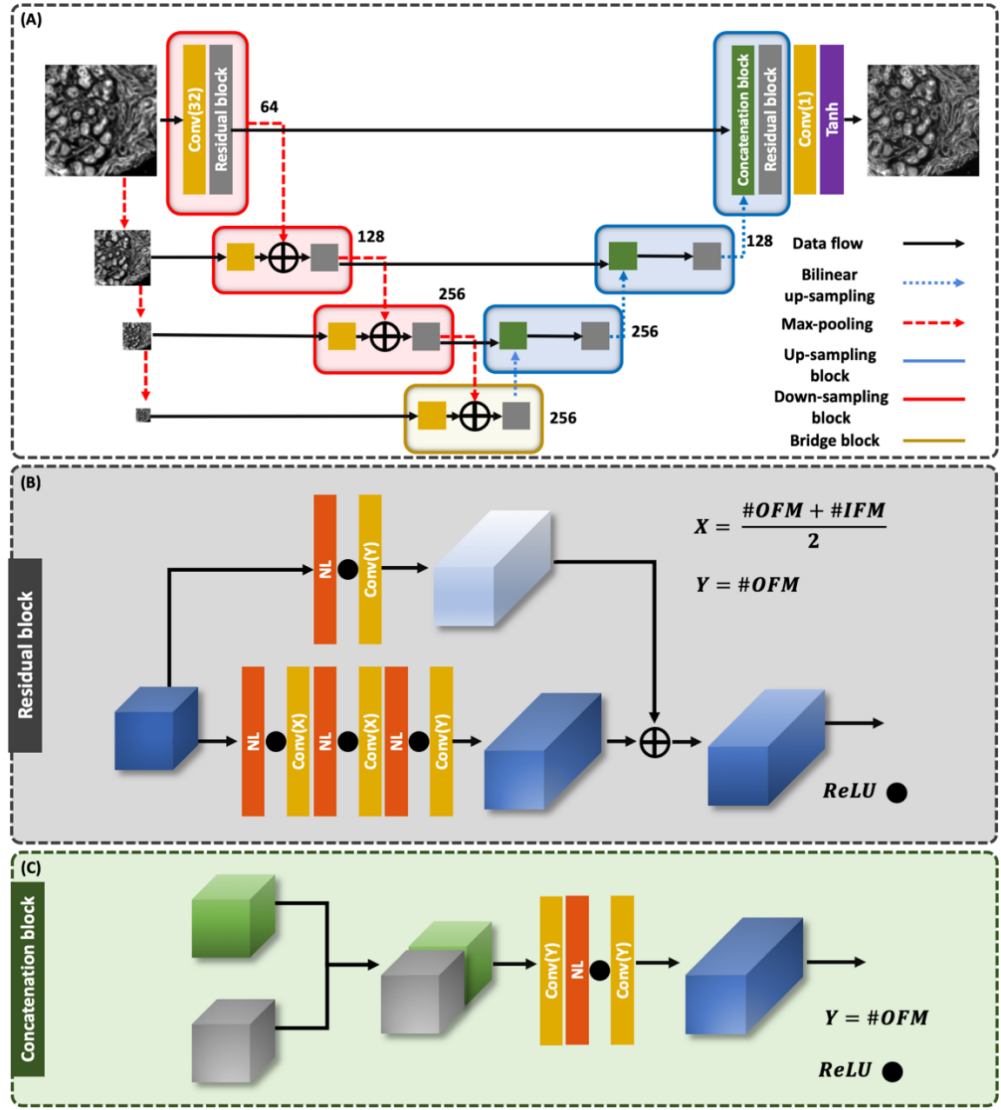

Fig. 2. The generator architecture (A) consists a series of down-sampling, bridge, and up-sampling blocks where the number of output feature maps (#OFM) is shown following each block. The CNN architectures representing the layouts of the (B) residual blocks and (C) concatenation blocks consist of convolution layers (Conv) where the argument represents the output feature maps (#OFM) and number of input feature maps (#IFM) as well as normalization layers (NL).

The generator architecture is a combination of U-Net[54] and Res-Net[55] and is illustrated in Fig. 2A consisting of three down-sampling blocks, a bridge block, and three up-sampling blocks. The down-sampling block includes a convolution layer followed by a residual block described in Fig. 2B and a max pooling layer at the end. The bridge block consists of a convolution layer followed by a residual block, and then by a bilinear up-sampling block. The layout of an up-sampling block includes a concatenation block and a residual block. The layout of each concatenation block can be found in Fig. 2C. Finally, the output of the last convolution operation is followed by tangent hyperbolic activation function. All these convolution layers have a filter size of 3x3 pixels, a stride of 1 pixel, and zero padding.

The architecture of the discriminator is shown in Fig. 3 and follows a design previously reported[42], The input image goes to several convolution layers and batch normalization layers. The result of the last batch normalization layer goes to an average pooling. Next the output fed into two fully connected layers. Finally, the result of the last fully connected layer (FCL) goes into a sigmoidal activation function to obtain the probability map.

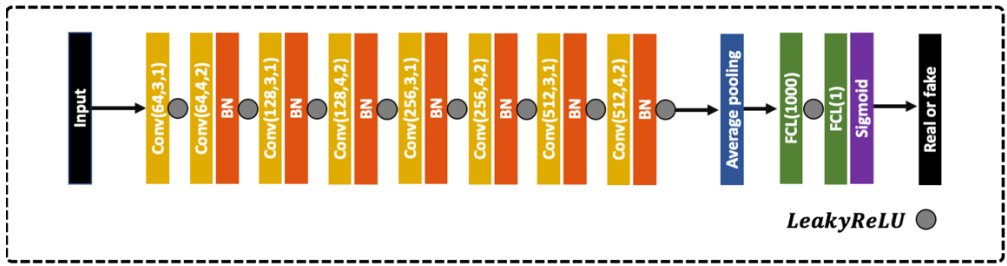

Fig. 3. The discriminator architecture consisting of convolution layers (Conv), batch normalization layers (BN), average pooling, and fully connected layers (FCL). The parameter of the leakyReLU is 0.2.

We implemented two normalization layers. For U-GAN, we implemented batch normalization[56]. In the C-GAN framework, we use a conditional normalization layer[52] to better incorporate texture features of tissue. The architecture of the conditional normalization layer used for the C-GAN is described by Fig. 4. The input activations go into a batch normalization layer[56] while the input class labels go into two different series of convolutional layers (convolutional layer, RELU and second convolutional layer). For each series, the number of feature maps resulting from the input class labels matches the number of feature maps resulting from batch normalization. The output of the first series is multiplied to the resulting batch normalization feature maps. Then this product is added to the output feature maps of the second series. Therefore, the normalization of every feature map is related to the specific input class labels. The input class labels to the network consist of one mask for each class in separate channels.

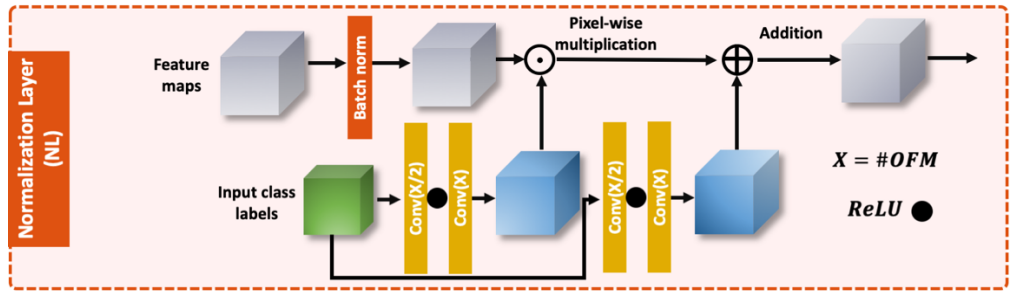

Fig. 4. The layout of normalization layer. X represent the number of output feature maps resulting of convolution layer. This conditional normalization layer scale and shift the output of batch normalization layer corresponding to each class label.

## 2.3 Loss functions

We first train the model's performance with a mean squared error (MSE) loss function calculated as,

$$\mathcal{L}_{MSE} = \frac{1}{R*C} \sum_{1}^{R} \sum_{1}^{C} (X_H - G(X_L))^2 \qquad (1)$$

where R and C are the row and column pixel dimensions of the image, while $X_H$ and $G(X_L)$ are the true high-resolution and generated high-resolution images respectively when given a low-resolution image $X_L$.

The MSE and all of the pixel-wise loss functions encourage the network to overly smooth and blur the outputs, thus this low perceptual quality making it impossible to increase the apparent resolution[42,57,58]. Fig. 5 shows the IR image and network output with only the MSE loss function illustrating that while the network can maintain the low frequency spatial information, it is unable to recover finer structures. To recover the lost high-frequency details, we utilize perceptual loss by using the feature maps of the last convolutional layer of a pre-trained VGG-19[42][59] and computed the MSE of the feature maps of the generated image and the real one as,

$$\mathcal{L}_{VGG} = \frac{1}{R'*C'} \sum_{1}^{R'} \sum_{1}^{C'} (VGG(X_H) - VGG(G(X_L)))^2 \qquad (2)$$

where R' and C' are the row and column pixel dimensions of the convolution layer. Note that the R' and C' are not equal to R and C.

We also incorporate an adversarial loss function where for the generative model is defined as,

$$\mathcal{L}_{adv,G} = -\log D(G(X_L)) \qquad (3)$$

and for the discriminative model, the function is defined as,

$$\mathcal{L}_{adv,D} = -\log D(X_H) - \log(1 - D(G(X_L))) \qquad (4)$$

Fig. 5C shows the effect of VGG features based loss and adversarial loss on the network output. It is evident that a VGG plus adversarial loss function perform much better as compared to a model based on a simple MSE loss function.

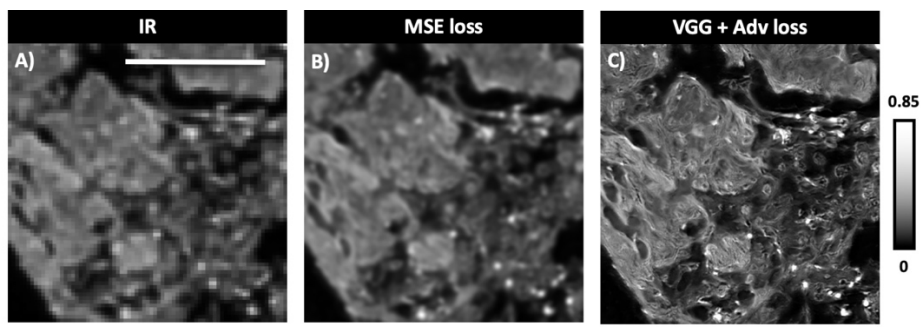

Fig. 5. Effect of different loss functions on the performance of network. A) LR IR (Network input). B) Network output with MSE loss. C) Network output with VGG features based loss and adversarial loss. The scale bar is 50 μm.

## 2.4 Training

We simulate the LR and HR datasets by transforming H&E images. Bright field microscopic H&E images are Gaussian blurred with a 3-pixel standard deviation and bilinearly down-sampled by a factor of 8 and then up-sampled back to the original size using nearest neighbor interpolation. This step is required to approximate the resolution of IR images. The simulated low-resolution image will serve as the input of the network and the true high-resolution image as the target. We randomly pick one of the RGB channels of the image for each iteration during training. In order to match the contrast of this grayscale H&E to IR images, we invert the gray scale H&E image. For data augmentation, besides the random rotation between 0 to 180 degrees, we change the contrast of the input and target by adjusting the exponent of the input and target images to a specific value between 0 and 4. We classify H&E images using VGG-19 pretrain model[59] and use the resulting images as class labels of the network in the training stage. Each patch has size of 96 x 96 and the batch size for each iteration is 12. All of images in the dataset are originally 256 x 256, for each iteration a random 96 x 96 region from the image is selected. The order of the images is random, and they are shuffled after each epoch. We used 5,600 images for training and 168 for validation. In the test stage, we use IR absorbance images at the Amide I (1658 cm$^{-1}$) band along with the classification data as the input. To scale the intensity of the IR images such that they match the network training data, we linearly normalize the absorbance values to the 90$^{th}$ percentile.

We trained two GANs[48]: U-GAN and C-GAN. Our generator G generates high resolution images given LR images as an input and our discriminator D critiques the generator via adversarial loss to improve the output. The instability of GANs in converging due to the adversarial formulation [60] necessitates a balance between G and D to prevent overfitting and outperforming either of them. Therefore, within each iteration of the discriminator, the generator iterates six times to avoid the discriminator from overfitting and outperforming the generator. Also, the learning rate of the discriminator is always 10% of the learning rate of the generator. The initialization of both the generative model and the discriminative model is random. Adam[61] was used to optimize the parameters of both G and D with an initial learning rate of $10^{-4}$ and $10^{-5}$ respectively. The training has two parts similar to the previous study[42]. First part is to make a pretrained model with only MSE loss function. To do so we train the model with 50,000 iterations with only MSE loss. The next part of training is using the pretrained model with 100,000 additional iterations as per the equations below:

$$\mathcal{L}_{\text{Total}} = \mathcal{L}_{\text{VGG}} * \alpha + \mathcal{L}_{\text{Adv,G}} * \gamma \quad (5)$$

The two regularization terms $\alpha$ and $\gamma$ in Eq. (5) stabilize the training to reach the convergence. With grid searching we determined that the optimum value of $\alpha$ and $\gamma$ is 0.01 and 0.005 respectively.

*2.5 Implementation details*

Computations were performed using the resources of the National Center for Supercomputing Applications at University of Illinois at Urbana-Champaign with a single NVIDIA V100 GPU NVDIA and Intel Xeon CPU E5-2680 v4 @ 2.4GHz. The framework is implemented in PyTorch 0.4.1, CUDA 9.2, and Python 3.7.1.

## 3. Results

First, we evaluate the performance of our network in enhance the morphological detail of a single band of IR image well beyond the limit of what can be recorded. As shown in Fig. 6, our model takes an input IR image and the corresponding classification data as semantic labels and generates a super-resolved image that is comparable to the ground truth image (H&E stained) of the adjacent section. Since the ground truth is from an adjacent tissue slice, it does not exactly match the IR image. In addition, we trained the same model but without any conditions and prior knowledge of object identity. Due to the similarities in low spatial frequencies details, recovering high spatial frequencies is challenging and error prone. Arrows show regions where the U-GAN model failed to increase the resolution correctly, even while it enhances spatial detail. Those areas are epithelial cells and the U-GAN model predicts it as a stromal structure instead. The major reason for the failure of the U-GAN is the hallucination problem. Deep learning models are susceptible to hallucination due to the insufficient or deficient dataset that can cause plausible artifacts if contradictory data does not exist[62]. However, by incorporating the class labels and using C-GAN, the deep neural network alleviates this problem and are able to recover the high spatial frequencies data with higher accuracy. This is crucial in the field of biomedical imaging where inaccuracy in the analysis can lead to misdiagnosis. In general, however, we note that the approach works well and can rapidly provide enhanced images with insignificant additional computational overhead (~ 1 s for a 1 MPx image). This speed is critical as rapidly generated SR images can be coupled with an IR imaging instrument to generate light microscope resolution-comparable images in real time. The generated label free HR images will approximate the spatial quality of light microscopy and the molecular contrast of a single band IR spectroscopic image. The model can be applied to every band of the IR data to generate the high resolution full spectral images. Due to the variation in the spatial details and contrast within each band, the predicted morphology is not the same which may lead to inaccuracy in the spectral profiles at the pixel level.

Before the use of deep learning in the field of microscopy, deconvolution methods were commonly used to increase the spatial quality. Deconvolution typically provides a 2-4-fold improvement in spatial detail, with a model for the image formation optics. Our deep learning framework enables us to produce super-resolved images directly, given images from diffraction limited imaging system such as IR imaging. Unlike deconvolution methods, our network improves the spatial resolution without any information of imaging configurations, understanding of the PSF of the imaging system, or need of any manual estimation of parameters [63,64]. However, knowing the actual PSF helps to simulate better the low resolution images, thereby, increasing the quality of generated images. Another essential component of our super resolution network is data augmentation (see the methods), which allows us to transform the spatial information from different imaging modalities by conserving the histogram. This allows us to directly simulate the high-resolution image, from the low resolution one, without any prior knowledge of another form of super-resolved IR imagery.

Such image data sets, to our knowledge, do not exist. Thus, the method proposed here is not a means to obtain better data, faster or with less effort; it is a means to estimate data that is otherwise simply not available. One other advantage of using the simulated training dataset of our method is that there is no need to precisely register low resolution images to high resolution images, which sometimes is impossible due to sample corruption and manipulation during imaging.

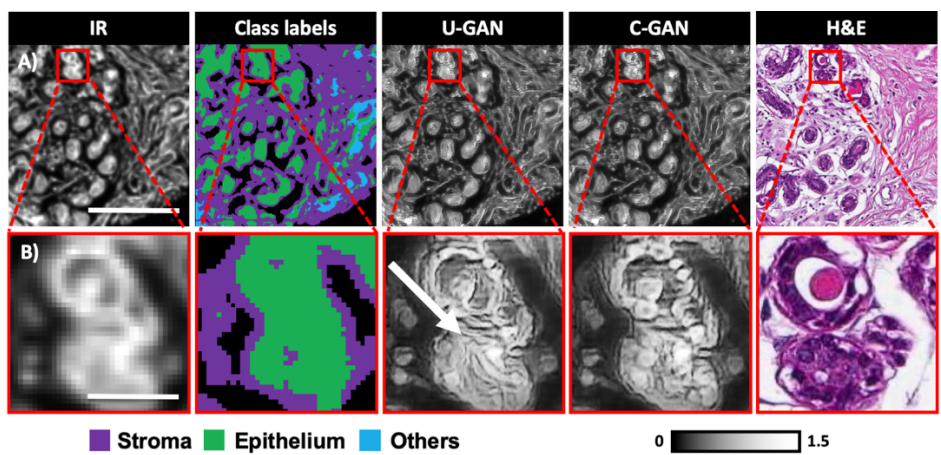

Fig. 6. Visual comparison of network's output for C-GAN and U-GAN. (A) is a breast sample and (B) is the zoomed in region. From left to right we show IR image, class labels, U-GAN's output, C-GAN's output, and the adjacent H&E. The scale bar for the (A) is 100 μm and for (B) is 35 μm. The arrow shows a sample region where U-GAN estimate the texture incorrectly.

Obviously, as can been seen in Fig. 6, the method works well but on the limited pathologies and tissue architectures that are common in breast tissue. In order to apply our framework to other types of tissue, training on the corresponding tissue's dataset is recommended to get optimal results. Different types of samples have different cell types, morphologies, and structures which introduce new challenges if they are unseen to the network at the training stage. Using transfer learning, a model that has been trained previously on a different tissue type, can expedite the convergence of the network using fewer number of iterations, or it can help the learning process when we are limited by the size of the dataset[65]. A generalized approach to obtaining image details, and validating the approach, is obviously desirable but difficult. We note that our purpose was to improve image detail to help decision-making in pathology. In this use case scenario, the application of the developed approach is highly justified as there is a set of defined morphological structures and their specific alterations in disease. Further, there is a large volume of cases for a particular organ/disease and tailored algorithms tuned for specific tissues/pathologies can be easily deployed. There are several methods to incorporate modifications to the deep learning network to account for specific use cases and extend the utility of the developed approach. class labels can be concatenated to the LR image as an input to the network; however, it has been shown [53] that this approach may not recover high frequency spatial details accurately. Another approach is to decompose the images based on each class label. Then, train deep learning models for each class separately and combine the results of each model for every class. This approach is very computationally expensive and, when the number of classes increases, it becomes impractical. Regardless, the work presented here is a good starting point to explore such expansions.

One of the possible applications of this study is in all-digital histopathology. The traditional methodology of classification for IR imaging only takes into account spectral information[2,13] at every pixel. Recent work has shown that by taking into account both spatial and spectral information during training, for example, the classifier is capable of increased accuracy for automated diagnosis[66]. With the framework reported in this study and its capability for generating higher resolution data, the proposed methods can be coupled with classifiers that leverage both spectral and spatial details to perform accurate tissue segmentation tasks quickly. While enhancing the diagnostic capability of chemical imaging systems, integration with gold standard images used in the clinic can allow for smooth transition and acceptance. However, care needs to be exercised and extensive validation conducted to ensure robustness. To robustly achieve a high level of accuracy, training must be carefully guided by an appropriate loss function. We use three different loss functions to train our network: MSE loss, perceptual loss, and adversarial loss. The MSE loss function is a pixel-wise loss function that is necessary to use as a baseline to guide the network during the initial optimization iterations, however by only minimizing this loss function, the network tends to generate estimations that are excessively smoothed[42,53]. To overcome this challenge, similar to previous studies[42,43,53], we incorporate perceptual loss and adversarial loss functions. By using perceptual and adversarial loss functions, we encourage the network to generate images which does not necessarily match the ground truth in a pixel-wise manner, but will have similar textural representations of the actual object. The combination of all of these loss functions are recommended in order to recover both low-level and high-level features, encouraging the network to estimate the morphology of the sample in IR domain beyond the diffraction limit.

As with any other machine learning algorithm, our model has some possible failure cases. First, lack of sufficient contrast between histologic units makes it difficult for the deep neural network to successfully generate correct HR images. For instance, when nuclei are very close to each other, resolution is difficult and correct predictions are not feasible. Another reason for failures may arise from a variety of sub-classes within each class. As illustrated in Fig. 7, for example, "stroma" is easy for pathologists to recognize but contains different sub-types including dense, loose and reactive. Each of these types has a different morphology and contrast, for instance reactive stroma consists of an immune cells and fibroblasts, while dense and loose stroma are mainly made up of extracellular materials and fibers with a markedly lower density of cells. Obviously, a model of great complexity and ever-increasing detail can be used. Here, our goal was to focus on epithelial cells since they are the predominant cell type involved in tumors and their morphology has diagnostic significance. While the role of the stroma in cancer progression is acknowledged, in contrast, there are no broadly accepted morphologic markers. Hence, we used a simplified network here by training on only one "stroma" class. This leads to some deficiencies. For example, the area indicated by the red arrow in Fig. 7 shows a reactive stromal region due to the accumulation of lymphocytes. However, the serial section H&E suggests a dense stroma for that area. There are two possible explanations for this aberration. First, since we are comparing enhanced-resolution IR with a serial section H&E, there might be different extent of inflammation in the two sections. Therefore, there is a possibility that the model prediction is correct, but it is somewhat unlikely. Secondly, this accumulation of lymphocytes in the C-GAN output can be caused by an unusual level of scattering in the IR input image. This unusual level of scattering can arise from an unusual arrangement of fibers, local stromal breakdown, processing variations, surface roughness or leftover paraffin in the stroma. Thus, while the prediction is likely incorrect in this case, a highly scattering stroma from an IR image should and indeed is recognized as reactive stroma.

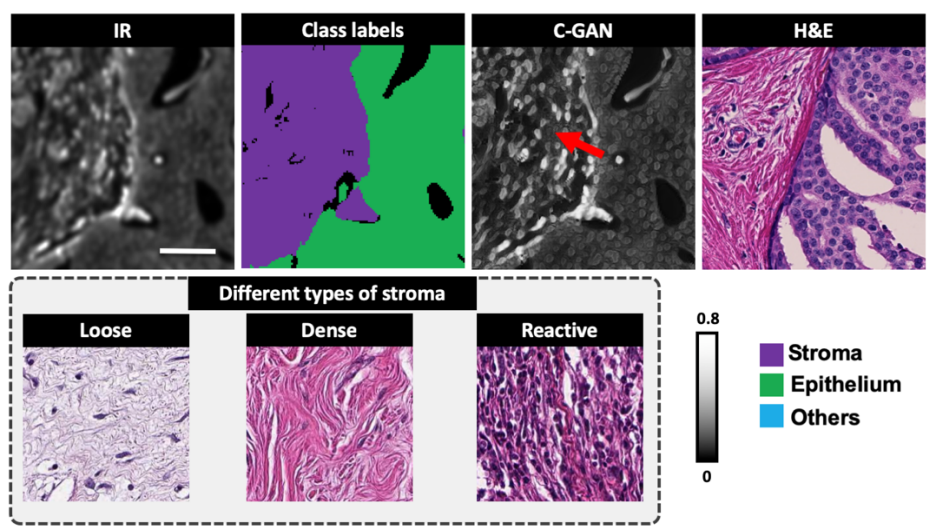

Fig. 7. Illustration of a potential failure case. The top row from left to right is the input IR image, class labels, output image of the C-GAN, and a serial section's H&E stained visible microscopy image. The bottom row shows different types of stroma encountered in breast tissue for comparison. The scale bar is 30 μm.

## 4. Conclusion

We report a deep learning framework that increases the morphologic detail of diffraction limited IR chemical images to approximate those obtained at a resolution defined by light microscopy. This framework represents a means to exceed performance beyond the capability of modern instrumentation in the field of IR spectroscopy, allowing both new morphologic detail and retention of the exceptional molecular imaging content of spectroscopic imaging. Our approach is based on a simple inclusion of the typical images used for diagnostic pathology and, consequently, is constrained to provide details that are diagnostically relevant. Unlike deconvolution methods[63,64], our frameworks rapidly performs resolution enhancement, without the need for any additional parameters and optimization. Therefore, it can be easily integrated to any IR microscope for real time acquisition offering ease of deployment and use. The proposed approach can further be extended to any imaging modality thereby making it useful for a wide range of applications. This work is generalized for an arbitrary IR microscope, however, by taking into account the measured response of a specific system, we can improve the low-resolution image estimations. This will enhance the quality of the resulting calculated SR data. This work can open new opportunities for obtaining high resolution IR images with intact spectral fidelity.

## 5. Funding, acknowledgments, and disclosures

*5.1 Funding*

This work is supported by the National Institutes of Health through grants R01CA197516 and R01EB009745.

*5.2 Acknowledgments*

This work utilizes resources provided by the Innovative Systems Laboratory at the National Center for Supercomputing Applications at the University of Illinois at Urbana-Champaign.

*5.3 Disclosures*

The authors declare no conflicts of interest.

**References**

1. I. W. Levin and R. Bhargava, "Fourier transform infrared vibrational spectroscopic imaging: integrating microscopy and molecular recognition," Annu. Rev. Phys. Chem. **56**, 429–474 (2005).
2. D. C. Fernandez, R. Bhargava, S. M. Hewitt, and I. W. Levin, "Infrared spectroscopic imaging for histopathologic recognition," Nat. Biotechnol. **23**(4), 469 (2005).
3. S. G. Kazarian and K. L. A. Chan, "Applications of ATR-FTIR spectroscopic imaging to biomedical samples," Biochim. Biophys. Acta (BBA)-Biomembranes **1758**(7), 858–867 (2006).
4. M. N. Gurcan and A. Madabhushi, "Digital Pathology," in *Proc. of SPIE Vol* (2016), **9791**, p. 979101.
5. T. Wrobel and R. Bhargava, "Infrared spectroscopic imaging advances as an analytical technology for biomedical sciences," Anal. Chem. (2017).
6. S. Pahlow, K. Weber, J. Popp, R. W. Bayden, K. Kochan, A. Rüther, D. Perez-Guaita, P. Heraud, N. Stone, A. Dudgeon, and others, "Application of vibrational spectroscopy and imaging to point-of-care medicine: A review," Appl. Spectrosc. **72**(101), 52–84 (2018).
7. D. Finlayson, C. Rinaldi, and M. J. Baker, "Is infrared spectroscopy ready for the clinic?," Anal. Chem. **91**(19), 12117–12128 (2019).
8. M. Pilling and P. Gardner, "Fundamental developments in infrared spectroscopic imaging for biomedical applications," Chem. Soc. Rev. **45**(7), 1935–1957 (2016).
9. D. Mayerich, M. J. Walsh, A. Kadjacsy-Balla, P. S. Ray, S. M. Hewitt, and R. Bhargava, "Stain-less staining for computed histopathology," Technology **3**(01), 27–31 (2015).
10. P. Lasch and D. Naumann, "Spatial resolution in infrared microspectroscopic imaging of tissues," Biochim. Biophys. Acta (BBA)-Biomembranes **1758**(7), 814–829 (2006).
11. A. K. Kodali, M. Schulmerich, J. Ip, G. Yen, B. T. Cunningham, and R. Bhargava, "Narrowband midinfrared reflectance filters using guided mode resonance," Anal. Chem. **82**(13), 5697–5706 (2010).
12. M. R. Kole, R. K. Reddy, M. V Schulmerich, M. K. Gelber, and R. Bhargava, "Discrete frequency infrared microspectroscopy and imaging with a tunable quantum cascade laser," Anal. Chem. **84**(23), 10366–10372 (2012).
13. S. Mittal, K. Yeh, L. S. Leslie, S. Kenkel, A. Kajdacsy-Balla, and R. Bhargava, "Simultaneous cancer and tumor microenvironment subtyping using confocal infrared microscopy for all-digital molecular histopathology," Proc. Natl. Acad. Sci. 201719551 (2018).
14. K. Yeh, D. Lee, and R. Bhargava, "Multi-color Discrete Frequency Infrared Spectroscopic Imaging," Anal. Chem. (2019).
15. K. Yeh and R. Bhargava, "Discrete frequency infrared imaging using quantum cascade lasers for biological tissue analysis," in *Biomedical Vibrational Spectroscopy 2016: Advances in Research and Industry* (2016), **9704**, p. 970406.
16. R. K. Reddy, M. J. Walsh, M. V Schulmerich, P. S. Carney, and R. Bhargava, "High-definition infrared spectroscopic imaging," Appl. Spectrosc. **67**(1), 93–105 (2013).
17. M. J. Nasse, M. J. Walsh, E. C. Mattson, R. Reininger, A. Kajdacsy-Balla, V. Macias, R. Bhargava, and C. J. Hirschmugl, "High-resolution Fourier-transform infrared chemical imaging with multiple synchrotron beams," Nat. Methods **8**(5), 413 (2011).
18. L. S. Leslie, T. P. Wrobel, D. Mayerich, S. Bindra, R. Emmadi, and R. Bhargava, "High definition infrared spectroscopic imaging for lymph node histopathology," PLoS One **10**(6), e0127238 (2015).
19. R. Furstenberg, C. A. Kendziora, M. R. Papantonakis, V. Nguyen, and R. A. McGill, "Chemical imaging using infrared photothermal microspectroscopy," in *Next-Generation Spectroscopic Technologies V* (2012), **8374**, p. 837411.
20. A. Mërtiri, T. Jeys, V. Liberman, M. K. Hong, J. Mertz, H. Altug, and S. Erramilli, "Mid-infrared photothermal heterodyne spectroscopy in a liquid crystal using a quantum cascade laser," Appl. Phys. Lett. **101**(4), 44101 (2012).
21. A. Mertiri, H. Altug, M. K. Hong, P. Mehta, J. Mertz, L. D. Ziegler, and S. Erramilli, "Nonlinear midinfrared photothermal spectroscopy using Zharov splitting and quantum cascade lasers," ACS photonics **1**(8), 696–702 (2014).
22. D. Zhang, C. Li, C. Zhang, M. N. Slipchenko, G. Eakins, and J.-X. Cheng, "Depth-resolved mid-infrared photothermal imaging of living cells and organisms with submicrometer spatial resolution," Sci. Adv. **2**(9), e1600521 (2016).
23. Z. Li, K. Aleshire, M. Kuno, and G. V Hartland, "Super-resolution far-field infrared imaging by photothermal heterodyne imaging," J. Phys. Chem. B **121**(37), 8838–8846 (2017).
24. A. Centrone, "Infrared imaging and spectroscopy beyond the diffraction limit," Annu. Rev. Anal. Chem. **8**, 101–126 (2015).
25. S. Kenkel, A. Mittal, S. Mittal, and R. Bhargava, "Probe--Sample Interaction-Independent Atomic Force Microscopy--Infrared Spectroscopy: Toward Robust Nanoscale Compositional Mapping," Anal. Chem. **90**(15), 8845–8855 (2018).
26. E. C. Mattson, M. J. Nasse, M. Rak, K. M. Gough, and C. J. Hirschmugl, "Restoration and spectral

recovery of mid-infrared chemical images," Anal. Chem. **84**(14), 6173–6180 (2012).
27. R. Bhargava, S.-Q. Wang, and J. L. Koenig, "Processing FT-IR imaging data for morphology visualization," Appl. Spectrosc. **54**(11), 1690–1706 (2000).
28. C. Ni, Q. Li, and L. Z. Xia, "A novel method of infrared image denoising and edge enhancement," Signal Processing **88**(6), 1606–1614 (2008).
29. R. Bhargava, T. Ribar, and J. L. Koenig, "Towards faster FT-IR imaging by reducing noise," Appl. Spectrosc. **53**(11), 1313–1322 (1999).
30. S. N. Pandya, B. J. Peterson, K. Mukai, R. Sano, A. Enokuchi, and N. Takeyama, "Improved signal to noise ratio and sensitivity of an infrared imaging video bolometer on large helical device by using an infrared periscope," Rev. Sci. Instrum. **85**(7), 73107 (2014).
31. K. I. Willig, B. Harke, R. Medda, and S. W. Hell, "STED microscopy with continuous wave beams," Nat. Methods **4**(11), 915 (2007).
32. S. Manley, J. M. Gillette, G. H. Patterson, H. Shroff, H. F. Hess, E. Betzig, and J. Lippincott-Schwartz, "High-density mapping of single-molecule trajectories with photoactivated localization microscopy," Nat. Methods **5**(2), 155 (2008).
33. T. Van Dijk, D. Mayerich, R. Bhargava, and P. S. Carney, "Rapid spectral-domain localization," Opt. Express **21**(10), 12822–12830 (2013).
34. H. Hou and H. Andrews, "Cubic splines for image interpolation and digital filtering," IEEE Trans. Acoust. **26**(6), 508–517 (1978).
35. P. Thévenaz, T. Blu, and M. Unser, "Image interpolation and resampling," Handb. Med. imaging, Process. Anal. **1**(1), 393–420 (2000).
36. L. Zhang and X. Wu, "An edge-guided image interpolation algorithm via directional filtering and data fusion," IEEE Trans. Image Process. **15**(8), 2226–2238 (2006).
37. J. Yang and T. Huang, "Image super-resolution: Historical overview and future challenges," Super-resolution imaging 20–34 (2010).
38. M. Ben-Ezra, Z. Lin, and B. Wilburn, "Penrose pixels super-resolution in the detector layout domain," in *2007 IEEE 11th International Conference on Computer Vision* (2007), pp. 1–8.
39. Z. Lin and H.-Y. Shum, "Fundamental limits of reconstruction-based superresolution algorithms under local translation," IEEE Trans. Pattern Anal. Mach. Intell. **26**(1), 83–97 (2004).
40. J. Kim, J. Kwon Lee, and K. Mu Lee, "Accurate image super-resolution using very deep convolutional networks," in *Proceedings of the IEEE Conference on Computer Vision and Pattern Recognition* (2016), pp. 1646–1654.
41. C. Dong, C. C. Loy, K. He, and X. Tang, "Learning a deep convolutional network for image super-resolution," in *European Conference on Computer Vision* (2014), pp. 184–199.
42. C. Ledig, L. Theis, F. Huszár, J. Caballero, A. Cunningham, A. Acosta, A. Aitken, A. Tejani, J. Totz, Z. Wang, and others, "Photo-realistic single image super-resolution using a generative adversarial network," in *Proceedings of the IEEE Conference on Computer Vision and Pattern Recognition* (2017), pp. 4681–4690.
43. X. Wang, K. Yu, S. Wu, J. Gu, Y. Liu, C. Dong, Y. Qiao, and C. C. Loy, "Esrgan: Enhanced super-resolution generative adversarial networks," in *European Conference on Computer Vision* (2018), pp. 63–79.
44. H. Wang, Y. Rivenson, Y. Jin, Z. Wei, R. Gao, H. Günayd\in, L. A. Bentolila, C. Kural, and A. Ozcan, "Deep learning enables cross-modality super-resolution in fluorescence microscopy," Nat. Methods **16**, 103–110 (2019).
45. E. Nehme, L. E. Weiss, T. Michaeli, and Y. Shechtman, "Deep-STORM: super-resolution single-molecule microscopy by deep learning," Optica **5**(4), 458–464 (2018).
46. Y. Rivenson, Z. Göröcs, H. Günaydin, Y. Zhang, H. Wang, and A. Ozcan, "Deep learning microscopy," Optica **4**(11), 1437–1443 (2017).
47. H. Byeon, T. Go, and S. J. Lee, "Deep learning-based digital in-line holographic microscopy for high resolution with extended field of view," Opt. Laser Technol. **113**, 77–86 (2019).
48. I. Goodfellow, J. Pouget-Abadie, M. Mirza, B. Xu, D. Warde-Farley, S. Ozair, A. Courville, and Y. Bengio, "Generative adversarial nets," in *Advances in Neural Information Processing Systems* (2014), pp. 2672–2680.
49. P. Isola, J.-Y. Zhu, T. Zhou, and A. A. Efros, "Image-to-image translation with conditional adversarial networks," in *Proceedings of the IEEE Conference on Computer Vision and Pattern Recognition* (2017), pp. 1125–1134.
50. T.-C. Wang, M.-Y. Liu, J.-Y. Zhu, A. Tao, J. Kautz, and B. Catanzaro, "High-resolution image synthesis and semantic manipulation with conditional gans," in *Proceedings of the IEEE Conference on Computer Vision and Pattern Recognition* (2018), pp. 8798–8807.
51. H. Zhang, I. Goodfellow, D. Metaxas, and A. Odena, "Self-attention generative adversarial networks," arXiv Prepr. arXiv1805.08318 (2018).
52. T. Park, M.-Y. Liu, T.-C. Wang, and J.-Y. Zhu, "Semantic Image Synthesis with Spatially-Adaptive Normalization," arXiv Prepr. arXiv1903.07291 (2019).
53. X. Wang, K. Yu, C. Dong, and C. Change Loy, "Recovering realistic texture in image super-resolution by deep spatial feature transform," in *Proceedings of the IEEE Conference on Computer Vision and Pattern*

*Recognition* (2018), pp. 606–615.
54. O. Ronneberger, P. Fischer, and T. Brox, "U-net: Convolutional networks for biomedical image segmentation," in *International Conference on Medical Image Computing and Computer-Assisted Intervention* (2015), pp. 234–241.
55. K. He, X. Zhang, S. Ren, and J. Sun, "Deep residual learning for image recognition," in *Proceedings of the IEEE Conference on Computer Vision and Pattern Recognition* (2016), pp. 770–778.
56. S. Ioffe and C. Szegedy, "Batch normalization: Accelerating deep network training by reducing internal covariate shift," arXiv Prepr. arXiv1502.03167 (2015).
57. M. Mathieu, C. Couprie, and Y. LeCun, "Deep multi-scale video prediction beyond mean square error," arXiv Prepr. arXiv1511.05440 (2015).
58. J. Johnson, A. Alahi, and L. Fei-Fei, "Perceptual losses for real-time style transfer and super-resolution," in *European Conference on Computer Vision* (2016), pp. 694–711.
59. K. Simonyan and A. Zisserman, "Very deep convolutional networks for large-scale image recognition," arXiv Prepr. arXiv1409.1556 (2014).
60. N. Kodali, J. Abernethy, J. Hays, and Z. Kira, "On convergence and stability of gans," arXiv Prepr. arXiv1705.07215 (2017).
61. D. P. Kingma and J. Ba, "Adam: A method for stochastic optimization," arXiv Prepr. arXiv1412.6980 (2014).
62. C. Belthangady and L. A. Royer, "Applications, promises, and pitfalls of deep learning for fluorescence image reconstruction," Nat. Methods 1 (2019).
63. F. Aguet, D. Van De Ville, and M. Unser, "Model-based 2.5-D deconvolution for extended depth of field in brightfield microscopy," IEEE Trans. Image Process. **17**(7), 1144–1153 (2008).
64. W. H. Richardson, "Bayesian-based iterative method of image restoration," JOSA **62**(1), 55–59 (1972).
65. D. S. Kermany, M. Goldbaum, W. Cai, C. C. S. Valentim, H. Liang, S. L. Baxter, A. McKeown, G. Yang, X. Wu, F. Yan, and others, "Identifying medical diagnoses and treatable diseases by image-based deep learning," Cell **172**(5), 1122–1131 (2018).
66. S. Berisha, M. Lotfollahi, J. Jahanipour, I. Gurcan, M. Walsh, R. Bhargava, H. Van Nguyen, and D. Mayerich, "Deep learning for FTIR histology: leveraging spatial and spectral features with convolutional neural networks," Analyst **144**(5), 1642–1653 (2019).